\documentclass[aps,prl,twocolumn,preprintnumbers,floatfix,nofootinbib,superscriptaddress,longbibliography]{revtex4}
\usepackage{amsfonts} 
\usepackage{amssymb} 
\usepackage{amsmath} 
\usepackage{graphicx} 
\usepackage{subfigure} 
\usepackage{array} 
\usepackage{dcolumn} 
\usepackage{latexsym} 
\usepackage{longtable} 
\usepackage{bbold}
\usepackage{xcolor}
\usepackage{multirow}
\usepackage{rotating}
\usepackage{comment}
\usepackage{zref-xr}
\usepackage{zref-user}
\usepackage{slashed}
\usepackage{ulem}

\newcommand{\bea}{\begin{eqnarray}}
\newcommand{\eea}{\end{eqnarray}}
\newcommand{\be}{\begin{equation}}
\newcommand{\ee}{\end{equation}}
\newcommand{\boldsigma}{\mbox{\boldmath $\sigma$}}
\newcommand{\bt}{\beta}

\newcommand{\al}{\alpha}

\newcommand{\bma}{\begin{pmatrix}}
\newcommand{\ema}{\end{pmatrix}}

\usepackage{textgreek}
\usepackage[unicode]{hyperref} 
\hypersetup{
    colorlinks=true,       
    linkcolor=blue,          
    citecolor=blue,        
    filecolor=blue,      
    urlcolor=blue           
}

\definecolor{green}{rgb}{0.1, 0.8, 0.1}

\usepackage{textgreek}
\usepackage[unicode]{hyperref} 

\begin{document}


\title{Neutrinoless double-beta decay in the neutrino-extended Standard Model}

\author{Wouter~Dekens}
\email{wdekens@uw.edu}
\affiliation{Institute for Nuclear Theory, University of Washington, Seattle WA 98195-1550, USA}

\author{Jordy~de~Vries}
\email{j.devries4@uva.nl}
\affiliation{Institute for Theoretical Physics Amsterdam and Delta Institute for Theoretical Physics,
University of Amsterdam, Science Park 904, 1098 XH Amsterdam, The Netherlands}
\affiliation{Nikhef, Theory Group, Science Park 105, 1098 XG, Amsterdam, The Netherlands}

\author{Emanuele~Mereghetti}
\email{emereghetti@lanl.gov}
\affiliation{Los Alamos National Laboratory, Theoretical Division T-2, Los Alamos, NM 87545, USA}

\author{Javier~Men\'endez}
\email{menendez@fqa.ub.edu}
\affiliation{Departament de F\'isica Qu\`antica i Astrof\'isica, Universitat de Barcelona, 08028, Spain}
\affiliation{Institut de Ci\`encies del Cosmos, Universitat de Barcelona, 08028, Spain}

\author{Pablo~Soriano}
\email{pablo-sf@hotmail.com}
\affiliation{Departament de F\'isica Qu\`antica i Astrof\'isica, Universitat de Barcelona, 08028, Spain}
\affiliation{Institut de Ci\`encies del Cosmos, Universitat de Barcelona, 08028, Spain}

\author{Guanghui Zhou}
\email{ghzhou@itp.ac.cn }
\affiliation{ CAS Key Laboratory of Theoretical Physics, Institute of Theoretical Physics,
	Chinese Academy of Sciences, Beijing 100190, P. R. China}
\affiliation{Institute for Theoretical Physics Amsterdam and Delta Institute for Theoretical Physics,
University of Amsterdam, Science Park 904, 1098 XH Amsterdam, The Netherlands}
\affiliation{Nikhef, Theory Group, Science Park 105, 1098 XG, Amsterdam, The Netherlands}

\preprint{LA-UR-23-22287}
\preprint{INT-PUB-23-009}

\begin{abstract}
We investigate neutrinoless double-beta decay ($0\nu\beta\beta$) in the minimal extension of the standard model of particle physics, the $\nu$SM, where gauge-singlet right-handed neutrinos give rise to Dirac and Majorana neutrino mass terms. We focus on the associated sterile neutrinos and argue that the usual evaluation of their contributions to $0\nu\beta\beta$, based on mass-dependent nuclear matrix elements, is missing important contributions from neutrinos with ultrasoft and hard momenta. We identify the hadronic and nuclear matrix elements that enter the new contributions, and calculate all relevant nuclear matrix elements for $^{136}$Xe using the nuclear shell model. Finally, we illustrate the impact on $0\nu\beta\beta$ rates in specific neutrino mass models and show that the new contributions significantly alter the $0\nu\beta\beta$ rate in most parts of the $\nu$SM parameter space.
\end{abstract}
\maketitle

{\it Introduction} ---
The standard model of particle physics (SM) in its original form \cite{Glashow:1959wxa,Salam:1959zz,Weinberg:1967tq} predicts massless neutrinos and is convincingly ruled out by neutrino oscillation experiments \cite{Workman:2022ynf}. A minimal extension of the SM, called the $\nu$SM, adds two or more right-handed neutrinos, $\nu_R$, which are singlets under the SM gauge groups and therefore called sterile neutrinos or, if their masses satisfy $m_{\nu_R}\gg  \mathcal O ({\rm eV})$, heavy neutral leptons  \cite{Abdullahi:2022jlv}. At the renormalizable level, apart from a kinetic term, sterile neutrinos have a Majorana and a Dirac mass term connecting them to the SM left-handed lepton doublet  and the Higgs field. Besides accomodating neutrino masses, this simple $\nu$SM has several intriguing features \cite{Asaka:2006ek,Shaposhnikov:2008pf,Canetti:2012kh,Drewes:2017zyw,Boyarsky:2018tvu,Dasgupta:2021ies}: 1) neutrinos generally become Majorana particles, leading to the violation of lepton number (LNV), 2) it is possible to account for the baryon asymmetry of the universe \cite{Davidson:2008bu}, 3) a very light sterile neutrino can be a dark matter candidate \cite{Drewes:2013gca,Kusenko:2009up,Drewes:2016upu,Boyarsky:2018tvu}.

Different experiments are sensitive to sterile neutrinos depending on their mass $M$.
For all mass ranges, however, neutrinoless double-beta decay (0$\nu\beta\beta$) plays a prominent role. 0$\nu\beta\beta$ is the most sensitive probe of LNV \cite{Agostini:2022zub}, with current limits on $0\nu\beta\beta$ half lives exceeding $10^{26}$ years \cite{KamLAND-Zen:2022tow,Agostini2020} and prospects for improvements by two orders of magnitude in the next decade \cite{Abgrall2021,Adhikari2022,Augier2022,Adams2021b,Albanese2021,DARWIN:2020jme,Adams:2022jwx}.  
For $M\gg \mathcal O (\mathrm{GeV})$, $0\nu\beta\beta$ decay is mainly driven by the exchange of light active neutrinos, 
and is proportional to the so-called effective neutrino mass $m_{\beta\beta}$.
For lighter $M$ there can be additional non-standard contributions from the exchange of sterile neutrinos that can enhance or suppress the $0\nu\beta\beta$ rates.

Contributions from sterile neutrinos to $0\nu\beta\beta$ have been studied extensively in the literature \cite{Blennow:2010th,Mitra:2011qr,Li:2011ss,deGouvea:2011zz,Faessler:2014kka,Barea:2015zfa,Giunti:2015kza,Asaka:2005pn,Asaka:2011pb,Asaka:2013jfa,Asaka:2016zib}. These works include the effect of the mass of the exchanged neutrinos by replacing the usual denominator of the massless neutrino propagator, $1/\mathbf k^{\,2}$, by a massive one, $1/(\mathbf k^{\,2} + M^2)$, in the LNV potential used in nuclear many-body calculations. In this Letter, we argue that this only captures one part of the $M$ dependence in $0\nu\beta\beta$ amplitudes and that consistent computations should include additional terms that can significantly alter $0\nu\beta\beta$ rate predictions.

{\it Lagrangian} ---
We consider a general setup with the SM Lagrangian supplemented by renormalizable interactions with $n$ gauge-singlet neutrino fields 
\begin{eqnarray}\label{eq:smeft}
\mathcal L &=&  \mathcal L_{SM} - \left[ \frac{1}{2} \bar \nu^c_{R} \, M_R \nu_{R} +\bar L \tilde H Y_\nu \nu_R + \rm{h.c.}\right]\,,
\end{eqnarray}
in terms of the lepton doublet $L=(\nu_L,\, e_L)^T$, while $\tilde H = i \tau_2 H^*$ with $H$ the Higgs doublet in the unitary gauge. $\nu_{R}$ is a column vector of $n$ right-handed sterile neutrinos, $Y_\nu$ a $3\times n$ matrix of Yukawa couplings and $ M_R$ a symmetric $n \times n$ matrix.  After electroweak symmetry breaking
 \bea\label{eq:mass}
 \mathcal L_m = -\frac{1}{2} \bar N^c M_\nu N +{\rm h.c.}\,,\qquad M_\nu = \bma 0 &M_D^*\\M_D^\dagger&M_R^\dagger \ema \,,
 \eea
where $N = (\nu_L,\, \nu_R^c)^T$, $M_D = \frac{v}{\sqrt{2}}Y_\nu^\dagger$ where $v\simeq 246$ GeV, and $M_\nu$ is a symmetric matrix diagonalized through \bea\label{Mdiag}
U^T M_\nu U  = {\rm diag}(m_1,\dots , m_{3+n})\,,\,\,\, N = U N_m\,,
\eea
where $U$ is the unitary neutrino mixing matrix, $m_i$ are real and positive and $\nu= N_m +N^c_m=\nu^c$. The active neutrino masses are light: $m_{1,2,3} \lesssim 1$ eV \cite{KATRIN:2021uub}. In the $\nu$SM
\bea \label{eq:cancel}
\sum_{i=1}^{n+3} U_{ei}^2 m_i = \left(M_\nu\right)_{ee}^* = 0\,,
\eea
which plays an important role for $0\nu\beta\beta$  \cite{deGouvea:2005er,Blennow:2010th}. 

{\it Active neutrinos} --- Recent years have seen the development of a chiral effective field theory $(\chi$EFT) derivation of the so-called neutrino potential that induces $nn \rightarrow pp +ee$ transitions. $\chi$EFT provides an expansion in $p/\Lambda_\chi$ where the scales are given by the pion mass or nuclear Fermi momentum, $p \sim m_\pi \sim k_F = \mathcal O(100\,\mathrm{MeV})$,  and the breakdown scale, $\Lambda_\chi \sim 4\pi F_\pi \sim m_N = \mathcal O(1\,\mathrm{GeV})$, in terms of the pion decay constant and the nucleon mass. The exchange of light virtual Majorana neutrinos 
gives rise to a leading-order (LO) contribution 
from neutrinos with momenta $|\boldsymbol k| \sim k_F$ and $k_0 \sim |\boldsymbol k|^2/\Lambda_\chi$, defined as potential contributions.

The order-by-order renormalizability of the $0\nu\beta\beta$ amplitude requires the promotion of an $nn \rightarrow pp +ee$ contact term to LO. This term captures contributions from hard neutrinos  with momenta $k_0 \sim |\boldsymbol k| \sim \Lambda_\chi$ \cite{Cirigliano:2018hja,Cirigliano:2019vdj} and has been recently included in many-body computations \cite{Wirth:2021pij,Weiss:2021rig}. The LO  $0\nu\beta\beta$ half-life reads
\bea\label{eq:rate}
\left(T_{1/2}^{0\nu}\right)^{-1}
	= G_{01}\,g_A^4  \Big |\sum_{i=1}^3 V_{ud}^2\frac{U_{ei}^2 m_i}{m_e}A_\nu \Big| ^2\,,
\eea
where $G_{01}$ is a phase-space factor  ($G_{01} = 1.4\cdot 10^{-14}$ y$^{-1}$ for ${}^{136}$Xe \cite{Neacsu:2015uja,Kotila:2012zza}),
\bea\label{eq:AnuPot}
A_\nu =  \frac{\mathcal M_F}{g_A^2}-\mathcal M_{GT}-\mathcal M_T-2 g_\nu^{NN} m_\pi^2\frac{\mathcal M_{F,sd}}{g_A^2}\,,
\eea
where the various $\mathcal M_{i}$ are nuclear matrix elements (NMEs),  see Ref.\ \cite{Agostini:2022zub} for an overview, $g_A \simeq 1.27$ is the nucleon axial charge, and $V_{ud} \simeq 0.97$. $ g^{NN}_\nu $ is a low-energy constant (LEC) associated with hard-neutrino exchange that in principle can be calculated with lattice QCD \cite{Cirigliano:2022rmf,Davoudi:2020gxs,Cirigliano2020May,Davoudi2022May}, but so far only phenomenological determinations are available \cite{Cirigliano:2021qko,Cirigliano:2020dmx,Richardson:2021xiu}. 

Other contributions appear at next-to-next-to-leading order (N${}^2$LO). Here we highlight a correction that plays an important role for sterile neutrinos. It arises from ultrasoft neutrinos with momentum scaling $k_0\sim |\mathbf k|\sim k_F^2/m_N$ \cite{Cirigliano:2017tvr}. The ultrasoft amplitude is given by
\bea\label{eq:ampusoftMS}
& & A_\nu^{\rm (usoft)}(m_i) = 
\frac{8 \pi R_A}{g_A^2}\sum_n \langle 0^+_f | \mathcal J_\mu | n\rangle\langle n| \mathcal J^\mu | 0^+_i\rangle \times \nonumber\\
& & \int \frac{d^{d-1}k}{(2\pi)^{d-1}} \frac{1}{E_\nu\left[E_\nu +\Delta E_1-i\epsilon\right]} +(\Delta E_{1}\to \Delta E_{2})\,,
\eea
where $E_\nu = \sqrt{\mathbf k^2+m_i^2} \simeq |\mathbf k|$, $\Delta E_{1,2} = E_{1,2}+E_n-E_i$, with $E_{i}$ and $E_{n}$ denoting the energies of the initial and intermediate states, and $E_{1,2}$ stand for electron energies and $R_A \simeq 1.2\,A^{1/3}\,\mathrm{fm}$. $\mathcal J_\mu$ is the single nucleon charged current evaluated between the initial $| 0^+_i\rangle$, final $| 0^+_f\rangle$ and a complete set of intermediate states $| n\rangle$.

{\it Sterile neutrinos} --- Previous works in the literature just combine
the neutrino potential for light active neutrinos with a mass-dependent one for sterile neutrinos:
\begin{equation}\label{eq:pot_massprop}
 \sum_{i=1}^3 \frac{ U^2_{ei}m_{i}}{\mathbf{k}^2 } \rightarrow  \sum_{i=1}^3 \frac{ U^2_{ei}m_{i}}{\mathbf{k}^2 } + \sum_{i=4}^{n+3} \frac{ U^2_{ei}m_{i}}{\mathbf{k}^2 + m_i^2 }\,,
\end{equation}
for all values of $m_i$. In practice, this is done by computing $\mathcal M(m_i)=-(\mathcal M_F/g_A^2-\mathcal M_{GT}-\mathcal M_T)(m_i)$ for a range of $m_i$ and then fitting to the functional form~\cite{Faessler:2014kka,Asaka:2016zib,Bolton:2019pcu,Fang:2021jfv,Bolton:2022tds} 
\begin{equation}\label{eq:naiveNME}
\mathcal M(m_i) =\mathcal M(0) \frac{\langle p^2\rangle}{\langle p^2\rangle +m_i^2}\,,
\end{equation} 
where the exact value of $\langle p^2 \rangle \sim k_F^2$ depends on the isotope and the applied nuclear many-body method. 

However, this approach does not include any other contributions which leads to several shortcomings
\begin{itemize}
\item  NMEs become ill-defined for $m_i  \geq \Lambda_\chi$ because the $\chi$EFT expansion does not converge for $m_i/\Lambda_\chi\gtrsim 1$. These sterile neutrinos must be integrated out at the quark level leading to local dimension-9 operators which, after evolution to low-energy scales, can be matched to $\chi$EFT \cite{Dekens:2020ttz}. The resulting LECs and NMEs cannot be obtained from Eq.~\eqref{eq:naiveNME} because the dimension-9 operator does not factorize. 
\item For $m_i \lesssim \Lambda_\chi$, Eq.~\eqref{eq:naiveNME} misses the LO contribution from hard neutrinos captured by the mass-dependent LEC, $g_\nu^{NN}(m_i)$. 
\item If for all sterile neutrinos $m_i \ll k_F$, the $0\nu\beta\beta$ rate is suppressed because of Eq.~\eqref{eq:cancel} \cite{deGouvea:2005er,Blennow:2010th}. In this limit, Eq.~\eqref{eq:naiveNME} predicts 
\begin{eqnarray}\label{suppression}
\left(T_{1/2}^{0\nu}\right)^{-1} = G_{01}\,g_A^4\Big |\sum_{i=1}^{n+3} V_{ud}^2\frac{U_{ei}^2 m_i^3}{m_e \langle p^2 \rangle}\mathcal M(0) \Big|^2\,,
\end{eqnarray} 
which is suppressed by $(m_i^2/\langle p^2\rangle)^2\sim m_i^4/k_F^4$. However, ultrasoft contributions suffer a milder suppression of $m^2_i/k_F^2$ and $(\frac{m_i^2}{4\pi \Delta E\,k_F}\, \ln \frac{m_i}{\Delta E} )^2$, where $\Delta E \sim k_F^2/m_N$ is a nuclear excitation energy. These effects lead to much faster decay rates. 
\end{itemize}
We now discuss an improved description of the $0\nu\beta\beta$ amplitude for different regions of $m_i$. 

\textit{Heavy masses: $m_i \geq \Lambda_\chi$}. Heavy sterile neutrinos can be integrated out at the quark level, giving rise to local operators containing four quarks and two electrons
\begin{equation}\label{eq:Ldim9}
\mathcal{L}^{(9)} = C_L (\mu_0) \bar{u}_L \gamma^\mu d_L \bar{u}_L \gamma_\mu d_L \bar{e}_L e^c_L\,,  
\end{equation}  
with $C_L (\mu_0) =-\eta(\mu_0,m_i)\frac{4 V^2_{ud} G^2_F}{m_i} U^2_{ei}$. $\eta(\mu_0,m_i)$ takes into account the QCD renormalization-group evolution from the scale $m_i$ to $\mu_0 =2$ GeV at which we match to $\chi$EFT. This evolution is mild and we include it into our results but we discuss it no further here. The matching to $\chi$EFT leads to hadronic LNV vertices \cite{Prezeau:2003xn,Graesser:2016bpz,Cirigliano:2018yza} and a resulting amplitude
\bea\label{A9}
A_\nu^{(9)} &=& -2\eta(\mu_0,m_i) \frac{m_\pi^2}{m_i^2}\bigg[\frac{5}{6}g_1^{\pi\pi}\bigg(\mathcal M_{GT,sd}^{PP} 
+\mathcal M_{T,sd}^{PP}\bigg)\nonumber\\
&& +\frac{g_1^{\pi N}}{2}\left(\mathcal M_{GT,sd}^{AP}+\mathcal M_{T,sd}^{AP}\right)
-\frac{2g_1^{NN}}{g_A^2}\mathcal M_{F,sd}\bigg],
\eea
where $g_1^{\pi\pi}$, $g_1^{N\pi}$, and $g_1^{NN}$ are hadronic LECs. The $\mathcal M_i$ denote NMEs \cite{Cirigliano:2018yza}, which have been calculated for several nuclei \cite{Hyvarinen:2015bda,Menendez:2017fdf,Barea:2015kwa}. In turn, $g_1^{\pi\pi}$ is currently the only LEC determined by lattice QCD~\cite{Nicholson:2018mwc,Detmold:2020jqv,Detmold:2022jwu}. Using the calculation of Ref.\ \cite{Nicholson:2018mwc}, we get
$g_{1}^{\pi\pi} = 0.36(2)$ at the scale $\mu =2$~GeV.
The naive limit of Eq.\ \eqref{eq:AnuPot} would yield the same expression as Eq.\ \eqref{A9}, but with $g_{1}^{\pi \pi} = 3/5$,
$g_{1}^{\pi N} = 1$, and $g_1^{NN} = (1 + 3 g_A^2)/4$. QCD effects thus cause 
$g_{1}^{\pi \pi}$ to differ by about a factor of 2 compared to the naive factorization approach. We expect similar deviations in $g_{1}^{\pi N}$
and $g_{1}^{NN}$, stressing the importance of controlling the hadronic input.

\textit{Intermediate masses: $k_F < m_i < \Lambda_\chi$}. In this mass region, sterile neutrinos appear as explicit degrees of freedom in $\chi$EFT \cite{Dekens:2020ttz}.  The potential contributions arise from Eq.~\eqref{eq:pot_massprop}, combined with the hard effects they give
\bea \label{eq:pot>}
A_\nu(m_i) =  -\mathcal M(m_i)-2 g_\nu^{NN}(m_i) m_\pi^2\frac{\mathcal M_{F,sd}}{g_A^2}\,,
\eea
which requires knowledge of the $m_i$-dependence of both the NMEs (from many-body calculations) and the LEC $g_\nu^{NN}$ (from non-perturbative QCD). In addition, there are contributions from loops involving soft sterile neutrinos, which for light neutrinos would contribute at N${}^2$LO but here can give rise to terms scaling as $m_i^2/\Lambda_\chi^2$. These effects lead to a breakdown of the $\chi$EFT expansion when $m_i \sim 1$ GeV. There are no contributions from ultrasoft sterile neutrinos because the integral in Eq.~\eqref{eq:ampusoftMS} vanishes in dimensional 
regularization once the integrand is expanded in  $k/m_i$.

\textit{Light masses: $ m_i < k_F$}. Here the potential and hard regimes are similar as for active neutrinos. A new effect appears due to ultrasoft sterile neutrinos. When $k_F>m_i$, we can perform the integrals of Eq.~\eqref{eq:ampusoftMS} 
\begin{eqnarray}\label{eq:ampusoftMSregion3}
A_{\nu}^{\rm (usoft)} &=& 
-\frac{ R_A}{2\pi}   \sum_{n}   \langle 0^+|  \tau^+ \boldsigma   |n\rangle \langle n|  \tau^+ \boldsigma  |0^+\rangle \nonumber\\
&&\times \left[  f(m_i, \Delta E_1)+ f(m_i, \Delta E_2)\right]   \,,
\end{eqnarray}
We give the detailed form of the loop function $f(m_i,\,\Delta E)$ in the appendix and focus on the leading $m_i$ dependence. For $m_i > \Delta E $, $f$ becomes independent of the energy splittings and $f\sim -\pi m_i$, while, for $m_i < \Delta E$,  $f\sim \frac{m_i^2}{\Delta E}(\frac{1}{2}+\ln \frac{2 \Delta E}{m_i})$. In both regions, the scaling with the sterile neutrino mass is more favorable than Eq.~\eqref{suppression}.

\textit{Nuclear and hadron matrix elements.} The correct description of the $0\nu\beta\beta$ amplitude depends on several new NMEs and LECs. We calculate all necessary NMEs using the nuclear shell model, one of the leading many-body methods used for $\beta\beta$ decay~\cite{Engel:2016xgb,Agostini:2022zub}. 
In the rest of this Letter, we focus on ${}^{136}$Xe but our conclusions apply to other experimentally relevant isotopes, such as $^{76}$Ge, as well. We use the GCN5082 effective Hamiltonian~\cite{Caurier:2010az} in a configuration space comprising the $0g_{7/2}$, $1d_{5/2}$, $2s_{1/2}$, $1d_{3/2}$ and $0h_{11/2}$ single-particle orbitals for protons and neutrons. We obtain our results with the shell-model code NATHAN~\cite{Caurier:2004gf}.
 
For potential contributions we evaluate the explicit $m_i$ dependence in the range $5\,\mathrm{MeV} < m_i < 2$ GeV, see the appendix. A fit to Eq.~\eqref{eq:naiveNME} gives $\langle p^2\rangle\simeq (175\, {\rm MeV})^2$. However, we use the functional form
\begin{equation}\label{eq:fitM}
\mathcal M(m_i) =\mathcal M(0)\frac{1}{1+m_i/m_a+(m_i/m_b)^2}\,,
\end{equation}
where $\mathcal M(0)=2.7$, $m_a =157$ MeV, and  $m_b = 221$ MeV  fit the calculated NMEs within a few-percent accuracy. Eq.~\eqref{eq:fitM} contains a linear term in $m_i$ different from the usually used functional form in Eq.~\eqref{eq:naiveNME}.

The ultrasoft contributions require the intermediate-state energies of ${}^{136}$Cs, $E_n$, in addition to matrix elements 
involving also the ${}^{136}$Xe and  ${}^{136}$Ba ground states. We use the Lanczos strength function method~\cite{Caurier:2004gf}, which after 60 iterations gives converged results for $\mathcal A_{\nu}^{\text{(usoft)}}$. Typical energy differences are $E_n-E_i~\sim1-10$~MeV, while the electron energies are $E_1\simeq E_2\simeq Q_{\bt\bt}/2+m_e $, with $Q_{\bt\bt}\simeq 2.5$ MeV for ${}^{136}$Xe, up to percent-level corrections of order $\mathcal O\left( (E_2-E_1)^2/(\Delta E_{1,2})^2 \right)$. All calculated NMEs are given in the appendix.

The hard contributions depend on a hadronic and a nuclear matrix element: $\mathcal M_{F,\mathrm{sd}} \times g_\nu^{NN}(m_i)$ which only in combination with $\mathcal M_{\mathrm{}}(m_i)$ is independent of the regulators used in nuclear computations \cite{Cirigliano:2018hja}. 
The value of $g_\nu^{NN}(m_i)$ thus depends on the nuclear many-body method used. We follow Refs.~\cite{Cirigliano:2019vdj,Richardson:2021xiu,Jokiniemi:2021qqv} and connect $g_{\nu}^{NN}(0)$ to charge-indepedence-breaking nucleon-nucleon interactions, in good agreement with model estimates \cite{Cirigliano:2020dmx,Cirigliano:2021qko}. From the nuclear shell model we get  $\mathcal M_{F,\mathrm{sd}} = -1.94$ for $g_{\nu}^{NN}(0) = -1.01\,\mathrm{fm}^2$  \cite{Jokiniemi:2021qqv}. The $m_i$-dependence is harder to pin down. Around $m_i \sim \Lambda_\chi$ the sum of the potential and hard contributions should match to Eq.~\eqref{A9} which requires $g_{\nu}^{NN}(m_i \sim \Lambda_\chi) \sim m_i^{-2}$. In the opposite limit,  $m_i\ll k_F$, we have $g_{\nu}^{NN}(m_i)\simeq g_{\nu}^{NN}(0)+g_{\nu,2}^{NN}m_i^2$ with $g_{\nu,2}^{NN}=\mathcal O(f_\pi^{-2}\Lambda_\chi^{-2})$ from $\chi$EFT power counting. These scalings are obeyed by the functional form 
\begin{equation}\label{eq:gnu_int}
g_{\nu}^{NN}(m_i) = g_{\nu}^{NN}(0) \frac{1+ (m_i/m_c)^2}{1 + (m_i/m_c)^2(m_i/m_d)^2}\,,
\end{equation}
where $m_c = \mathcal O(\Lambda_\chi)$. Setting $m_c =1 $ GeV for concreteness we fix $m_d$ by matching to Eq.~\eqref{A9} at $m_i = \mu_0$. To get a reasonable estimate we saturate $A_\nu^{(9)}$ with the $g_1^{\pi \pi}$ and $g_1^{NN}$ contributions and set $g_1^{NN}\simeq (1+3 g_A^2)/4$, the factorization estimate, and $g_1^{\pi \pi} = 0.36$ \cite{Nicholson:2018mwc}. This recipe gives $m_d \simeq 146$ MeV and provides the most uncertain part of our analysis. The uncertainties in this description can be reduced by lattice QCD computations of all LECs in Eq.~\eqref{A9}.

\textit{A practical formula.} Finally, we connect the improved amplitude for different regions of $m_i$ in the following way
\begin{equation}\label{eq:fullint}
\begin{aligned}
A_\nu (m_i) = \begin{cases}
A_\nu^{\rm (<)}+A_\nu^{\text{(usoft)}}\,, &m_i <  100 \text{ MeV} \\
A_\nu^{\rm }\,,&\text{0.1}\, \text{GeV} \le m_i <  2 \text{ GeV} \\
A_\nu^{\text{(9)}}\,, & {\rm }\,  2 \text{ GeV} \le m_i 
\end{cases}
\end{aligned}
\end{equation}
where $A_\nu^{\rm (<)}= A_\nu^{\rm }+ m_i \frac{d}{d m_i} {\cal M}(m_i)$.  
In the region $m_i < k_F$, Eq.~\eqref{eq:fitM} contains a linear term (after expanding) in $m_i$, not included in the standard functional form of Eq.~\eqref{eq:naiveNME}. This linear term is dominated by ultrasoft contributions and to avoid double counting between Eqs.\ \eqref{eq:fitM} and \eqref{eq:ampusoftMSregion3} we remove it in the definition of  $A_\nu^{\rm (<)}$. The appearance of this linear term in $\cal A_\nu^{\rm usoft}$ and $\cal M$ allows for a consistency check, which we discuss in more detail in future work.

{\it Phenomenology and specific models} ---  To illustrate the important effects of our findings, we consider two toy models that capture the essential features of realistic seesaw models with light sterile neutrinos. We give more details in the appendix.

{\it The 3+1 Scenario} ---   We begin with the 3+1 scenario with three light active neutrinos and one sterile neutrino. 
The $4\times 4$ mass matrix has the following form
\begin{align}
 M_\nu = \bma 0  & M^*_{D,i}\\ 
 M^*_{D,i} & |M_R|e^{i\al_R}		      
 \ema\,,
 \end{align}
and we set  $M_{D,1} = M_{D,2} = M_{D,3} \equiv |M_{D}| e^{i\al_D}$ for simplicity. The model leads to two massless neutrinos and is thus ruled out but illustrates the importance of ultrasoft corrections. Diagonalizing the mass matrix leads to one active neutrino mass, $m_3$, the sterile mass, $m_4$, and 
the  mixing angles 
\bea
U_{e3}^2=-\frac{m_4}{m_3}U_{e4}^2 =- \frac{1}{3}\frac{m_4}{m_3+m_4}e^{i(2\alpha_D+\alpha_R)}\,,
\label{eq:3plus1}
\eea 
where $\alpha_{D,R}$ drop out in the $0\nu\beta\beta$ rate. 

\begin{figure}[t!]
	\includegraphics[width=0.48\textwidth]{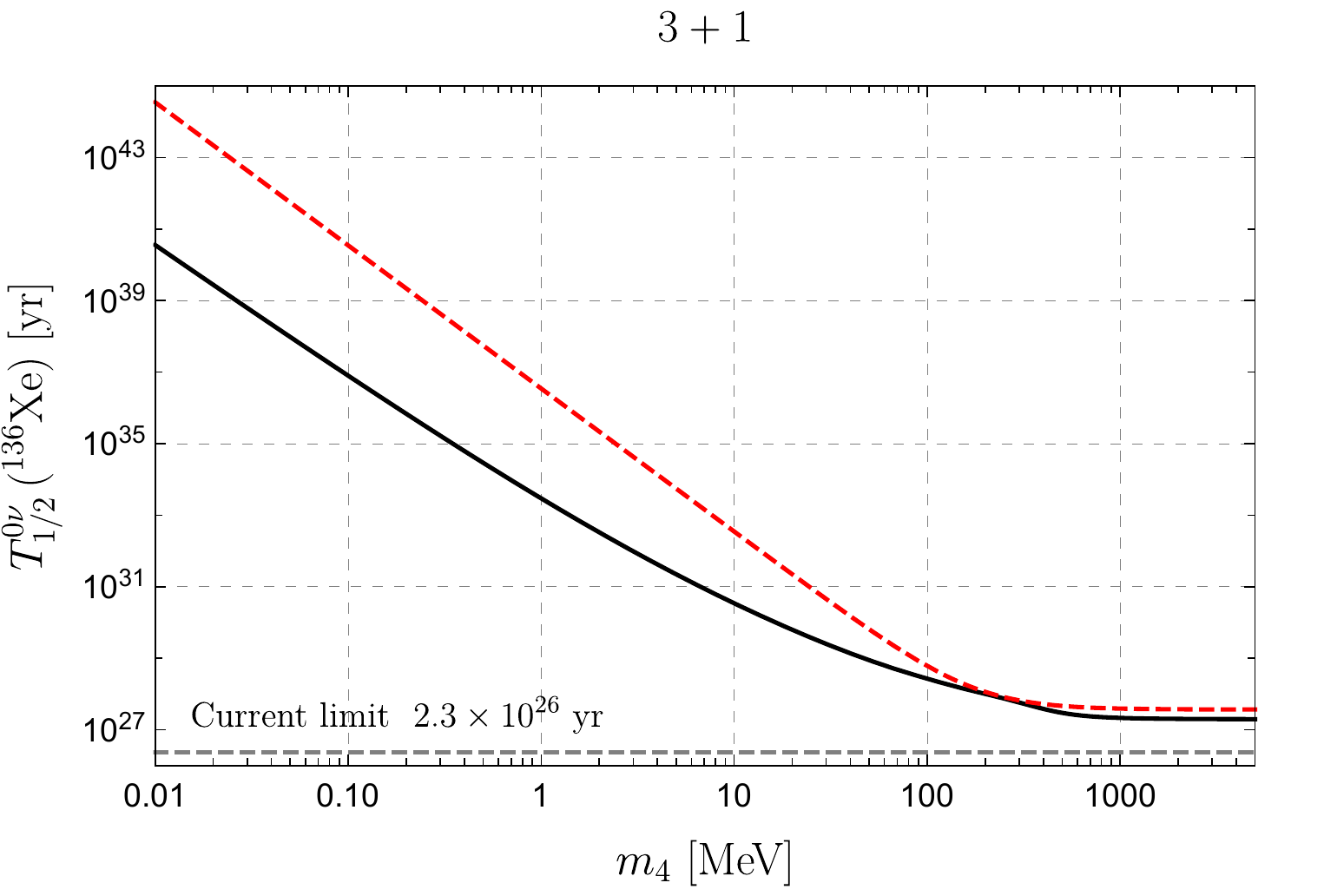}
	\includegraphics[width=0.48\textwidth]{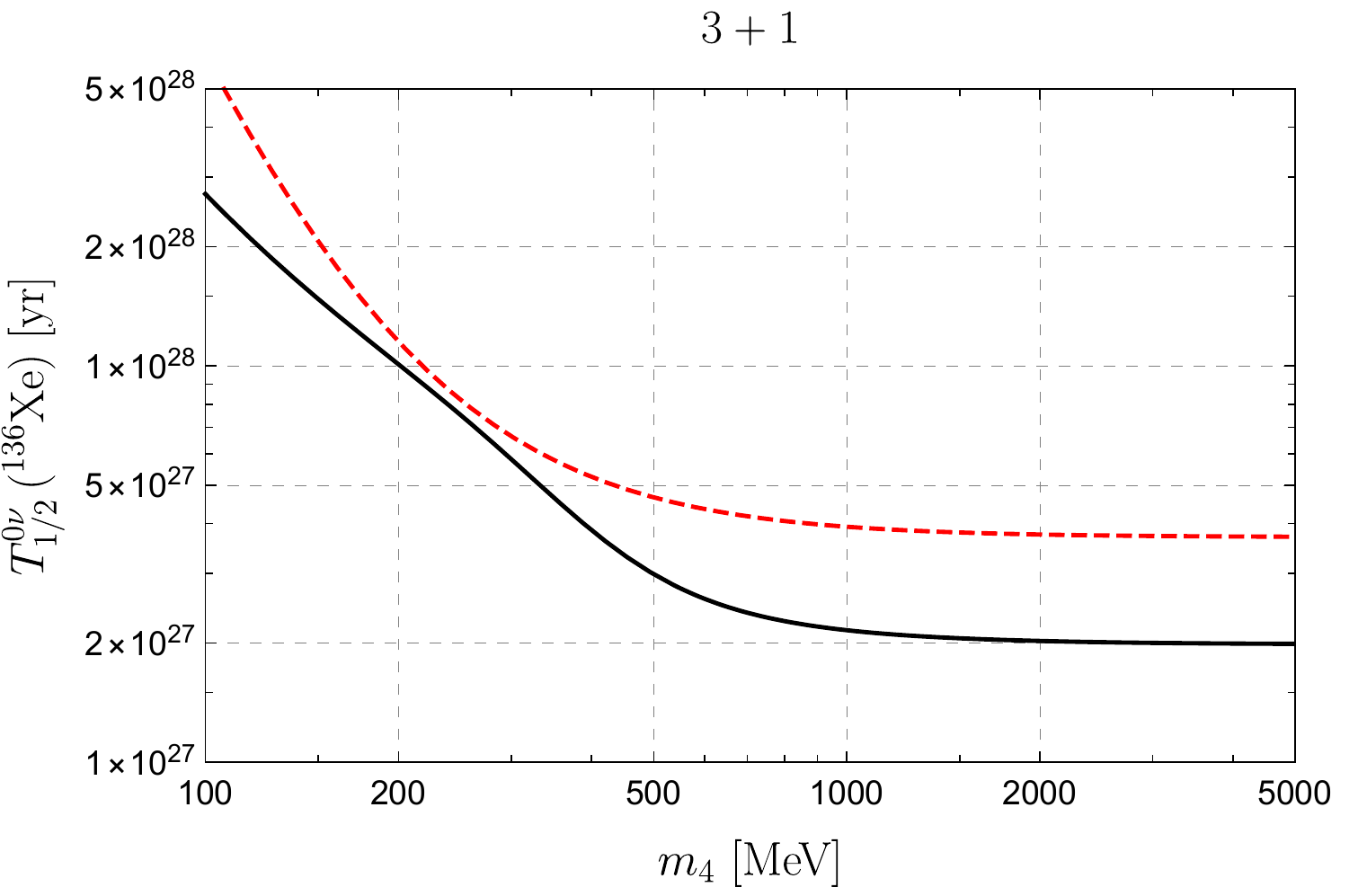}
	\caption{$0\nu\beta\beta$ half-life of $^{136}{\text{Xe}}$ as a function of $m_4$ in the 3+1 model, obtained with the NMEs in  Eq.\ \eqref{eq:naiveNME} (dashed red) and \eqref{eq:fullint} (black).  The bottom panel focuses on the results for larger $m_4$ values.}
	\label{fig:Thalf3p1}
\end{figure}

We set $m_3\simeq 0.05$ eV and show the resulting half-life of $^{136}{\text{Xe}}$ in Fig.\ \ref{fig:Thalf3p1}.  In the light $m_4$ regime, the lifetimes obtained from our approach (solid black) are much shorter than those obtained from Eq.\ \eqref{eq:naiveNME} (roughly three orders of magnitude for $m_4 = 10$ MeV) because of the ultrasoft contributions. The enhancement for $m_4 >$ 100 MeV is smaller, about a factor of 2, and mainly caused by hard-neutrino contributions. Such an enhancement is also found for light active neutrinos \cite{Cirigliano:2018hja,Wirth:2021pij,Jokiniemi:2021qqv}. 

{\it The Pseudo-Dirac Scenario} --- The masses of active neutrinos discussed above are inversely proportional to the Majorana mass of the sterile neutrino. Mechanisms exist in which  the active and sterile Majorana masses become proportional to a small LNV parameter, leading to pseudo-Dirac sterile neutrinos, see Refs.\ \cite{Dev:2012sg,BhupalDev:2012jvh,Bolton:2019pcu}. Here we consider the 
 inverse seesaw 1+2  scenario \cite{Mohapatra:1986bd,Mohapatra1986Feb,Nandi1986Feb} with 1 active and 2 sterile neutrinos. The $3 \times 3$ mass matrix has  the form
\bea\label{eq:inv_seesaw}
   M_\nu = 
   \bma
   0& m_D & 0\\
   m_D & \mu_X & m_S\\
   0 & m_S & \mu_S
   \ema\,.
   \eea
 Together with the assumption $m_S\gg m_D,\, \mu_{X,S}$, the mass of the lightest neutrino satisfies $m_\nu \simeq - (m_D^2 \mu_S)/m_S^2$
and is proportional to the small LNV parameter, $\mu_S$. We set $m_\nu = 2.6 \times 10^{-3}$ eV (as a typical value of $m_{\beta\beta}$ in the normal hierarchy) and write $M_{1,2}$ for the masses of the heavier neutrinos. We focus on a scenario where the heavier states act as a pseudo-Dirac pair with a small mass splitting (see the appendix). Variants of these models appear in scenarios of low-scale leptogenesis, see e.g. Refs.~\cite{Canetti:2012kh,Drewes:2021nqr, Hernandez:2022ivz}. 

\begin{figure}[t!]
	\center	
	\includegraphics[width=0.47\textwidth]{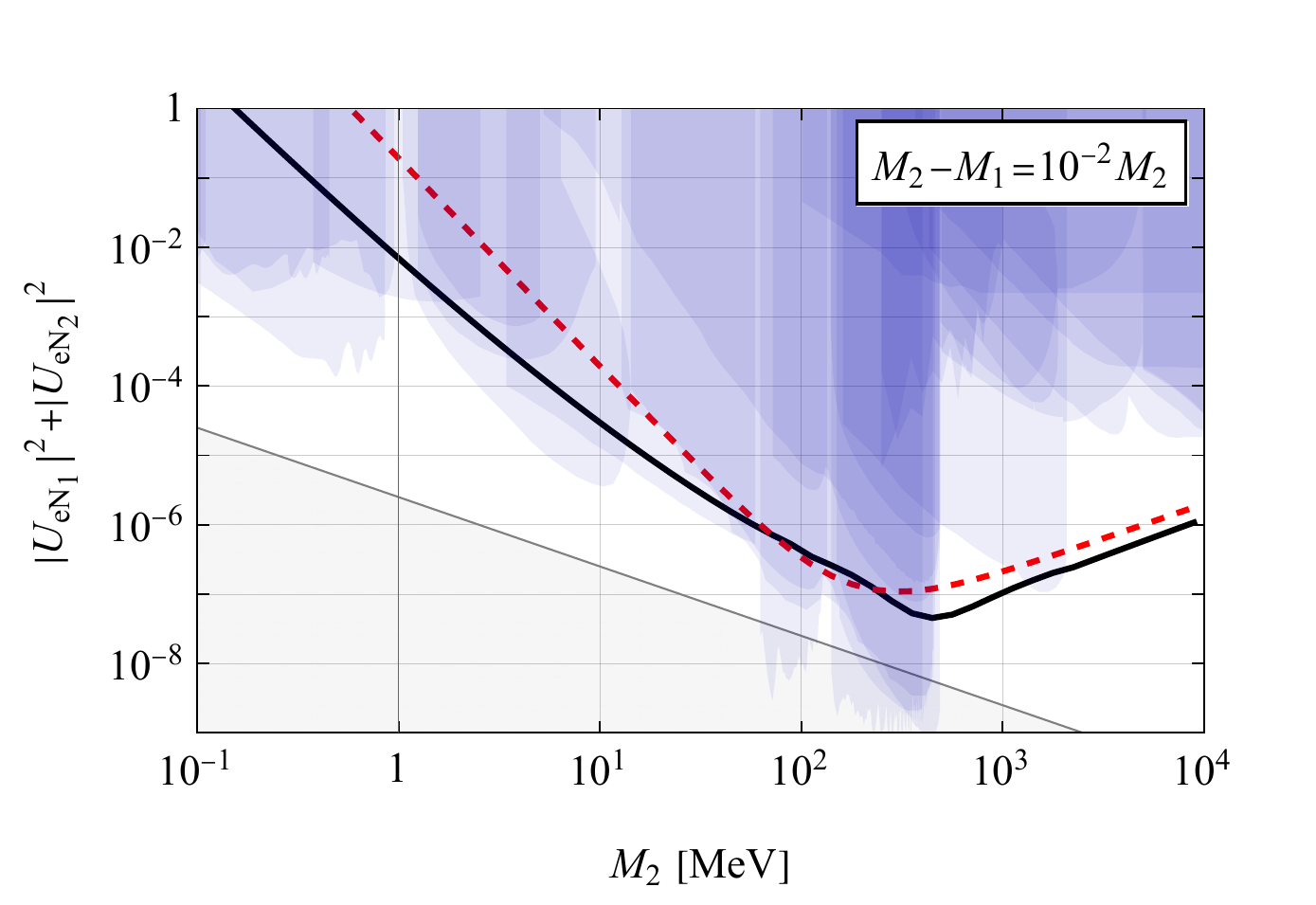}
	\caption{Limit on the mixing angles, $|U_{eN_1}|^2+|U_{eN_2}|^2$, as a function of $M_2$ in the pseudo-Dirac scenario for a mass splitting $\frac{M_2-M_1}{M_2}=10^{-2}$. The black line corresponds to the $0\nu\beta\beta$ rate obtained in this work, whereas the red dashed line is based on Eq.~\eqref{eq:naiveNME}. The blue background depict other experiments \cite{Bolton:2019pcu,bolton}, while the gray region indicates parameter space inconsistent within the assumptions of the model. }
	\label{fig:pseudo}
\end{figure}

The current limit on the half-life of $^{136}{\text{Xe}}$ \cite{KamLAND-Zen:2022tow} constrains the effective active-sterile mixing angle $|U_{eN_1}|^2+|U_{eN_2}|^2$ (note that $|U_{eN_1}|^2 \approx |U_{eN_2}|^2$ up to $\mathcal{O} (\Delta/M_2)$ corrections where $\Delta = M_2-M_1$). Fig.~\ref{fig:pseudo} shows these limits as a function of $M_2$ for $\Delta/M_2 = 1\%$ . Limits for smaller (larger) splittings weaken (strengthen) as $\Delta^{-1}$.  The black line denotes limits obtained by using the results of this Letter which are compared to literature approaches (through Eq.\ \eqref{eq:naiveNME}) shown in red. As in the $3+1$ model, the ultrasoft terms lead to significantly tighter constraints on the mixing angles for $M_2 < 100$ MeV. The $0\nu\beta\beta$ constraints are suppressed by the small mass splitting but nevertheless are competitive with other limits (indicated in purple and obtained from Refs.~\cite{Bolton:2019pcu,bolton}) for masses lighter than a few hundred MeV. For heavier masses, the hard neutrino-exchange contributions again lead to somewhat tighter limits (a factor 2.5 for $m_i \simeq 400$ MeV) than obtained from Eq.~\eqref{eq:naiveNME} and are significantly stronger than other constraints despite the small mass splitting.

\bigskip
{\it Conclusions and outlook ---} We have performed a systematic EFT derivation of the $0\nu\beta\beta$ rate in the neutrino-extended standard model, allowing us to identify and calculate novel contributions to  $0\nu\beta\beta$. The largest correction can enhance the decay rate by orders of magnitude and arises from the exchange of light, ultrasoft neutrinos which induce a sterile neutrino mass dependence of the $0\nu\beta\beta$ amplitude that differs from previous studies in the literature. We also find new contributions associated with the exchange of light, hard neutrinos, which can lead to differences compared to known expressions by a factor of a few. 
Thus, these new effects can significantly enhance the $0\nu\beta\beta$ rates in neutrino-mass models of interest. 

Looking to the future, our expressions can be made more accurate by first-principle determinations of various QCD matrix elements associated with virtual Majorana neutrino exchange. Lattice-QCD calculations of such matrix elements are already underway for active (and thus essentially massless) neutrino exchange \cite{Davoudi:2020gxs} and can be extended to massive neutrino exchange as well \cite{Tuo:2022hft}. The expressions obtained in this Letter can be directly used to determine $0\nu\beta\beta$ rates in realistic minimal neutrino extensions that can resolve shortcomings of the SM such as the neutrino-mass mechanism, leptogenesis, and dark matter.

\bigskip

\begin{acknowledgments}

\emph{Acknowledgements}.---
We thank Bhupal Dev for discussions on the results of Refs.\ \cite{Bolton:2019pcu,bolton}.
WD acknowledges support by the U.S. DOE under Grant No. DE-FG02-00ER41132. JdV acknowledges support from the Dutch Research Council (NWO) in the form of a VIDI grant. This work was supported by the ``Ram\'on y Cajal'' program with grant RYC-2017-22781, and grants CEX2019-000918-M and PID2020-118758GB-I00 funded by MCIN/AEI/10.13039/501100011033,  by ``ESF Investing in your future''.
EM was supported by the U.S. Department of Energy through the Los Alamos National Laboratory and by the Laboratory Directed Research and Development program of Los Alamos National Laboratory under project number 20230047DR.
Los Alamos National Laboratory is operated by Triad National Security, LLC, for the National Nuclear Security Administration of U.S. Department of Energy (Contract No. 89233218CNA000001). 

\end{acknowledgments}

\newpage
\newpage
\section{APPENDIX}
 
\begin{figure}[t]
\includegraphics[trim={1.5cm 0 0 1.9cm},clip ,width=0.45\textwidth]{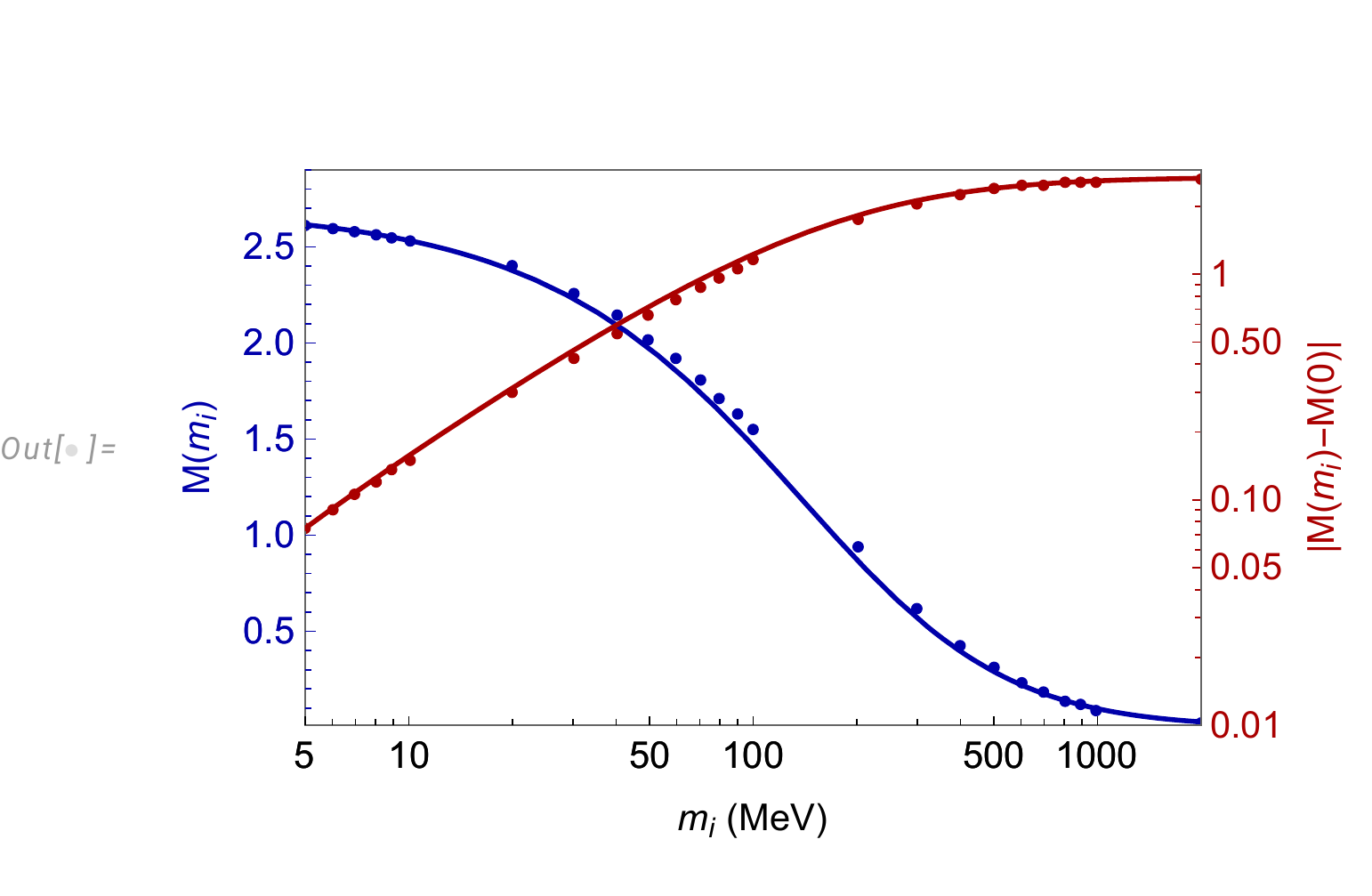}
\caption{NMEs in Eq.\ \eqref{eq:AnuPot} for $^{136}$Xe as a function of the neutrino mass (in blue), as well as the difference $\mathcal M(0)-\mathcal M(m_i)$ (in red). Circles indicate the numerical results of the shell-model calculations, while the solid lines depict the interpolation formula of Eq.\ \eqref{eq:fitM}.}
\label{fig:NMEs}
\end{figure}

{\it Nuclear matrix elements} --- Table \ref{tab:NME} lists the numerical values of the NME, $\mathcal M_\nu(m_i) = -(\mathcal M_F/g_A^2-\mathcal M_{GT}-\mathcal M_T)(m_i)$ of Eq.\ \eqref{eq:AnuPot}, as a function of the neutrino mass. Fig.\ \ref{fig:NMEs} shows this data, together with the interpolation of Eq.\ \eqref{eq:fitM}.

The computation of the ultrasoft contributions requires the values of the matrix elements and excited state energies in Eq.\ \eqref{eq:ampusoftMSregion3}, which are given in Table \ref{tab:overlapNME}. They must be combined with the function $f(m,E)$. For $m > E$ we have 
\begin{eqnarray}\label{eq:Fusoft}
f(m,E) &= &
-2\bigg[ E\left(1+ \ln \frac{\mu_{us}}{m} \right) +\sqrt{m^2 - E^2} \nonumber\\
&&\times  \left(\frac{\pi}{2}-\tan^{-1}\, \frac{ E}{\sqrt{m^2 -E^2}}\right) \bigg]\,,
\end{eqnarray}
while for $m <E$ 
\begin{eqnarray}
f(m,E) &=&  -2\bigg[ E \left(1+\ln \frac{\mu_{us}}{m} \right) \nonumber\\
&& -\sqrt{ E^2-m^2} \ln \frac{ E+\sqrt{ E^2-m^2}}{m}  \bigg]\,.
\end{eqnarray}
The leading $m_i$ dependence scales as $f\sim -\pi m$ for $m>E$, while for $m <E $, we have $f\sim E\frac{m^2}{E^2}(\frac{1}{2}+\ln \frac{2E}{m})$ as given in the main text. 
The dependence on the usoft renormalization scale, $\mu_{us}$, cancels in the amplitude when taking into account soft loop corrections to the potential \cite{Cirigliano:2017tvr}. We therefore set $\mu_{us} = m_\pi$, which captures the log-enhanced part of the soft loops, and stress that the exact choice of $\mu_{us}$ does not impact the main result which is the modified $m_i$ dependence of the amplitude.

 \begin{table}
    \begin{tabular}[b]{|c|ccccccccc|}    
    \hline
    $ m_i \,{\rm (MeV)}$
    &   5 & 
 6 & 
 7 & 
 8 & 
 9 & 
 10 & 
 20 & 
 30 & 
 40  \\\hline
     $M(m_i)$ &
  2.6 & 2.6  & 2.6  & 2.6  & 2.6  & 2.5  & 2.4  & 2.3  & 2.1   \\\hline
    $ m_i \,{\rm (MeV)}$ &50 &60
  &   70 & 
  80 & 
  90 & 
  100 & 
  200 & 
  300 & 
  400 \\\hline
  $M(m_i)$   & 2.0 &1.9  &
  1.8 & 1.7  & 1.6  & 1.5  & 0.94  &  0.61  & 0.42  \\\hline
     
  $ m_i \,{\rm (MeV)}$
 & 
500 & 
600 &700 &800 & 
  900 & 
  1000 & 
  2000 & 
   & 
    \\\hline
  $M(m_i)$  & 0.31  & 0.23 & 0.18 &0.14   & 0.11  &  0.094 & 0.025  &   &    \\\hline
  \end{tabular}
    \caption{Shell-model $0\nu\bt\bt$ NMEs for ${}^{136}$Xe as a function of the neutrino mass. }
    \label{tab:NME}
\end{table}

\begin{table*}[t]
	\renewcommand{\arraystretch}{1.3}    \centering
	\footnotesize
	\begin{tabular}{|c|c|c|}
		\hline
		$\frac{E_n-E_i}{\rm MeV}$&  $\langle n| \boldsigma\tau^+ | 0^+_i\rangle$& $\langle 0^+_f| \boldsigma\tau^+ | n\rangle$\\\hline
		0.17 & 1.0 & 0.13 \\
		0.63 & -0.19 & -0.0063 \\
		0.89 & -0.25 & -0.016 \\
		1.02 & 0.30 & 0.036 \\
		1.05 & 0.23 & 0.025 \\
		1.1 & -0.13 & -0.00076 \\
		1.2 & 0.12 & -0.0052 \\
		1.3 & 0.16 & -0.0028 \\
		1.4 & -0.23 & -0.0098 \\
		1.5 & 0.20 & -0.012 \\
		1.6 & -0.36 & 0.0084 \\
		1.7 & -0.24 & 0.00058 \\
		1.9 & 0.22 & 0.011 \\
		2.0 & 0.34 & 0.0070 \\
		2.2 & 0.35 & 0.0060 \\
		2.3 & -0.49 & -0.0086 \\
		2.6 & 0.62 & 0.021 \\
		2.7 & -0.91 & -0.024 \\
		2.9 & 0.37 & 0.0064 \\
		3.1 & 0.30 & 0.0013
		\\\hline
	\end{tabular}
	\qquad\qquad    
	\begin{tabular}{|c|c|c|}
		\hline
		$\frac{E_n-E_i}{\rm MeV}$&  $\langle n| \boldsigma\tau^+| 0^+_i\rangle$& $\langle 0^+_f| \boldsigma \tau^+ | n\rangle$\\\hline
		3.3 & 0.39 & -0.0013 \\
		3.6 & 0.39 & 0.0021 \\
		3.8 & 0.45 & -0.013 \\
		4.0 & -0.44 & -0.0032 \\
		4.3 & -0.35 & -0.0038 \\
		4.6 & -0.36 & -0.0067 \\
		4.8 & 0.44 & 0.0083 \\
		5.1 & 0.44 & 0.0066 \\
		5.4 & -0.55 & -0.0093 \\
		5.7 & 0.63 & 0.012 \\
		6.1 & 0.85 & 0.013 \\
		6.3 & -1.2 & -0.016 \\
		6.7 & -1.3 & -0.014 \\
		7.0 & -1.9 & -0.016 \\
		7.3 & 3.1 & 0.023 \\
		7.5 & -4.0 & -0.028 \\
		7.7 & 2.6 & 0.017 \\
		8.1 & 1.4 & 0.0091 \\
		8.4 & -1.0 & -0.0057 \\
		8.8 & -0.93 & -0.0064 
		\\\hline
	\end{tabular}
	\qquad\qquad
	\begin{tabular}{|c|c|c|}
		\hline
		$\frac{E_n-E_i}{\rm MeV}$&  $\langle n| \boldsigma\tau^+ | 0^+_i\rangle$& $\langle 0^+_f| \boldsigma\tau^+ | n\rangle$\\\hline
		9.1 & 0.80 & 0.0038 \\
		9.4 & 0.59 & 0.0014 \\
		9.8 & -0.50 & 0.0027 \\
		10.1 & 0.35 & -0.0027 \\
		10.5 & 0.26 & -0.00053 \\
		10.9 & -0.22 & -0.00021 \\
		11.3 & 0.17 & -0.00037 \\
		11.7 & -0.16 & -0.00054 \\
		12.0 & -0.16 & -0.0010 \\
		12.4 & 0.14 & 0.00092 \\
		12.8 & 0.12 & -0.00014 \\
		13.1 & 0.092 & -0.00040 \\
		13.5 & -0.079 & -0.00019 \\
		13.9 & 0.071 & -0.00026 \\
		14.2 & -0.070 & 0.000031 \\
		14.6 & -0.035 & 0.00021 \\
		15.1 & -0.051 & -0.00015 \\
		16.2 & -0.039 & 0.00011 \\
		17.3 & -0.043 & -0.000091 \\
		17.7 & 0.11 & -0.000029 
		\\\hline
	\end{tabular}    
	\caption{One-body matrix elements relevant for ultra-soft neutrino contributions to the $0\nu\bt\bt$ of $^{136}$Xe.}
	\label{tab:overlapNME}
\end{table*}

\bigskip
{\it The Pseudo-Dirac Model} --- 
The mass matrix in the Pseudo-Dirac model of Eq.\ \eqref{eq:inv_seesaw} can be diagonalized by
\bea 
U_{} = 
\bma 
1 & 0 & 0\\
0 & c_{12} & s_{12}\\
0 & -s_{12} & c_{12}\\ 
\ema \cdot
\bma 
c_{e2} & 0 & s_{e2}e^{-i\delta}\\
0 & 1 & 0\\
-s_{e2}e^{i\delta} & 0 & c_{e2}\\ 
\ema \\ \cdot
\bma 
c_{e1} & s_{e1} & 0\\
-s_{e1} & c_{e1} & 0\\ 
0 & 0 & 1
\ema\cdot
\bma 
1 & 0 & 0\\
0 & e^{i \alpha_1} & 0\\ 
0 & 0 & e^{i(\alpha_2+\delta)}
\ema\,,
\eea
where $s_{ij} = \sin\theta_{ij}$ and $c_{ij} = \cos\theta_{ij}$. 

Setting $\al_1=0$ and $\al_2=\pi/2$ allows the sterile neutrinos to act as a pseudo-Dirac pair. This choice, together with the constraint $\left(M_\nu\right)_{11}=0$, then allows one to write the $0\nu\bt\bt$ rate in terms of the neutrino masses and $|U_{eN_1}|^2+|U_{eN_2}|^2$. Combining this with $m_\nu = 2.6 \times 10^{-3}$ eV and $\Delta/M_2 = 10^{-2}$ then gives the lines in Fig.\ \ref{fig:pseudo}. The gray region depicts parts of parameter space where $3|\Delta|\leq|\left(M_\nu\right)_{22}|$, in tension with the pseudo-Dirac assumption, $\mu_X\sim\Delta$.
This gray region can be obtained by using the constraint $\left(M_\nu\right)_{13}=0$.

\bibliography{bibliography}

\end{document}